# Local Messages for Smartphones

Dmitry Namiot
Lomonosov Moscow State University
Faculty of Computational Mathematics and Cybernetics
Moscow, Russia
dnamiot@gmail.com

Manfred Sneps-Sneppe
Ventspils University College
Ventspils International Radio Astronomy Centre
Ventspils, Latvia
manfreds.sneps@gmail.com

*Abstract*— This paper describes a new model for local messaging based on the network proximity. We present a novelty mobile mashup which combines Wi-Fi proximity measurements with Cloud Messaging. Our mobile mashup combines passive monitoring for smart phones and cloud based messaging for mobile operational systems. Passive monitoring can determine the location of mobile subscribers (mobile phones, actually) without the active participation of mobile users. This paper describes how to combine the passive monitoring and notifications.

*Keywords— Wi-Fi;monitoring;proximity;cloud;messaging*

## I. INTRODUCTION

There are models of applications, where the concept of location can be replaced by that of proximity. At the first hand, this applies to use cases where the detection for exact location is difficult, even impossible or not economically viable [1]. Very often, this change is related to privacy. For example, a privacy-aware proximity detection service determines if two mobile users are close to each other without requiring them to disclose their exact locations [2]. As per developed algorithms for privacy-aware proximity detection methods we can mention papers [3] and [4], for example. They allow two online users to determine if they are close to each other without requiring them to disclose their exact locations to a service provider or other friends. Usually, the main goal for such systems is to generate proximity messages when friends approach each other closer than some predefined distance threshold. Technically, this threshold can be defined individually for each user (for group of users).

Of course, the term "distance" here depends on the metric used for the measurements. The classical example includes shortest path metric and two users on the different sides (banks) of the river. It is anti-pattern. The distance between users could be within the given threshold, but such "proximity" is useless.

Metric measurements for privacy can be replaced with some approximation by wireless proximity (network proximity). For this paper network proximity definition is very intuitive. It is a measure of how mobile nodes are close or far away from the elements of network infrastructure. There are several systems that can use network proximity as a base for mobile services. At the first hand, we can mention here our own system SpotEx (Spot Expert) [5]. According to this model, any existing or even especially created Wi-Fi hot spot could be used as presence sensor that can trigger access for some user-generated information snippets.

The typical application in this area uses collected database of so called Wi-Fi "fingerprints", including MAC addresses and the received signal strengths (RSSI) of nearby access points. This database could be used for Wi-Fi based positioning as well as for discovering the user's behavioral patterns [6]. A classical approach to Wi-Fi fingerprinting [7] involves RSSI (signal strength). The basic principles are transparent. At a given point, a mobile application may hear ("see") different access points with certain signal strengths. This set of access points and their associated signal strengths represents a label ("fingerprint") that is unique to that position. The metric that could be used for comparing various fingerprints is k-nearest-neighbors in signal space.

Problems associated with the collection of fingerprints, are fairly obvious. It is the price of the calibration process, the need for rework after the changes in the network and, most importantly, lack of support for dynamic networks. For example, most of the modern smart phones let users open Wi-Fi access point right on the phone.

We cannot create stable base of fingerprints for dynamic access points. Data linked to such dynamic access points becomes linked to the phones [1].

In our new service we've decided to use another fingerprints-less model: sniffing for beacon frames. It is a reverse schema for the standard fingerprints. We would like to analyze beacons transmitted by Wi-Fi devices, rather than beacons collected by them.

Collecting traces of Wi-Fi beacons is the well-know approach for getting the locations of Wi-Fi access points (AP). Beacon frames are used to announce the presence of a Wi-Fi network. As a result, an 802.11 client receives the beacons sent from all nearby access points. Client receives beacons even when it is not connected to any network. In fact, even when a client is connected to a specific AP, it periodically scans all the channels to receive beacons from other nearby APs. It lets clients keep track of networks in its vicinity. But in the same time Wi-Fi client periodically broadcasts an 802.11 probe request frame. Client expects to get back an appropriate probe request response from Wi-Fi access point. As per Wi-Fi spec, a station (client) sends a probe request frame when it needs to obtain information from another station. For example, a radio network interface card would send a probe request to determine which access points are within range. Figure 1 illustrates data flow for Probe Requests.

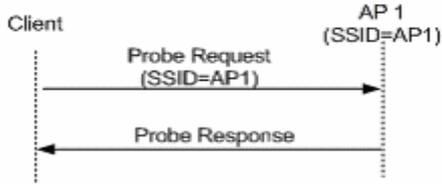

Fig. 1. Wi-Fi probe requests

A Probe Request frame contains two fields: the SSID and the rates supported by the mobile station. Stations that receive Probe Requests use the information to determine whether the mobile station can join the network. Drivers that allow cards to join any network use the broadcast SSID in Probe Requests [8]. Technically, probe request frame contains the following information:

- source address (MAC-address)
- SSID
- supported rates
- additional request information
- extended support rates
- vendor specific information

Our access point can analyze received probe request. Obviously, that any new request (any new MAC-address) corresponds to a new wireless customer nearby. Note, that Bluetooth devices could be monitored by the same principles.

Wi-Fi based device detection uses only a part from the above mentioned probe request. It is a device-unique address (MAC address). This unique information lets us re-identify devices (mobile phones) across our monitors.

We should note also, that passive Wi-Fi detection is not 100% reliable. Mobile phones (mobile OS, actually) can actually transmit probe requests at their discretion. Our own experiments with commercially available Wi-Fi probe scanners confirm data from [9]. Monitor detects in average about 70% of passing smartphones.

There are commercial off-the-shelf components that can provide passive Wi-Fi monitoring. For example, it is Meshlium Xtreme [10]. With passive monitoring Wi-Fi devices can be detected without the need of being connected to a specific access point, enabling the detection of any smartphone, laptop or hands-free device which comes into the coverage area.

Another component could be mentioned here is Cisco Mobility Services Engine (MSE). This equipment with Location Services provides presence detection and real-time location tracking, including track and trace of rogue devices, interferers, Wi-Fi clients, smart phones, and RFID tags [11]. Figure 2 is based on Cisco's manual and illustrates the simplified usage model.

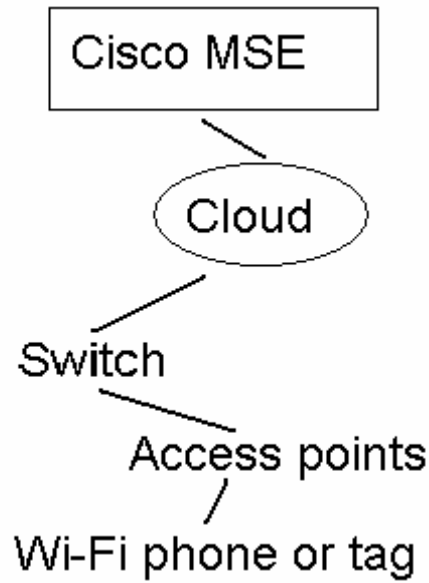

Fig. 2. Cisco MSE usage model

As the next entry in this category we can mention Navizon system [12]. Navizon I.T.S. is a system designed to track the location of WiFi stations in confined spaces. Examples of stations that can be tracked include smart phones, laptop computers, tablets, and any other device with a WiFi interface. Stations are tracked using special nodes that are deployed in the area to be monitored. Figure 3 illustrates this.

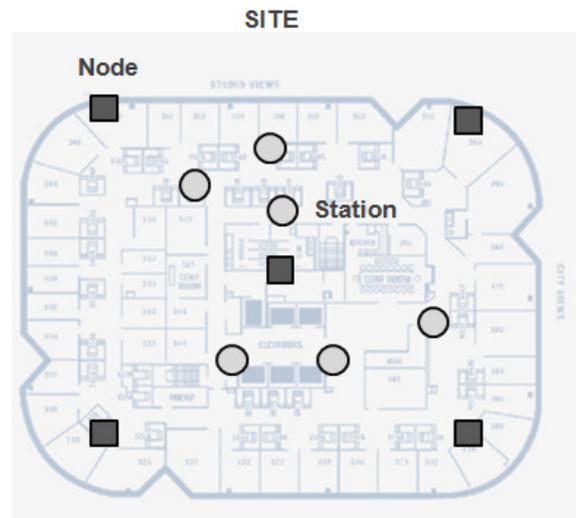

Fig.3. Navizon I.T.S.

Note, that the key moment in passive detection is MAC-address for mobile device. But in the same time, in statistical applications we will use it for re-identification only. It means, by the way, that for keeping the privacy we do not need to save in our database an original address. It is enough just to keep some hash-code for this address.

The typical tasks this approach could be applied for are:

- get a number of people passing daily in a street

- detect an average time of the stance of the people in a street or in a building

- differentiate between residents (daily matches) and visitants (sporadic matches)

- detect the walking routes of people in shopping malls and average time in each area

Technically, this so-called location analytics follows to the web sites analytics. Any new MAC-address registered by the scanning device is an analogue for web site hit. So, the main analytical report that is available in such system will show hits (unique MAC-adresses) per time. The set historical data lets us detect patterns as well as unusual behavior (e.g., this Friday is significant different comparing with historical data, etc.)

Figure 4 (by Cisco) illustrates Time of Day Distribution from location analytics.

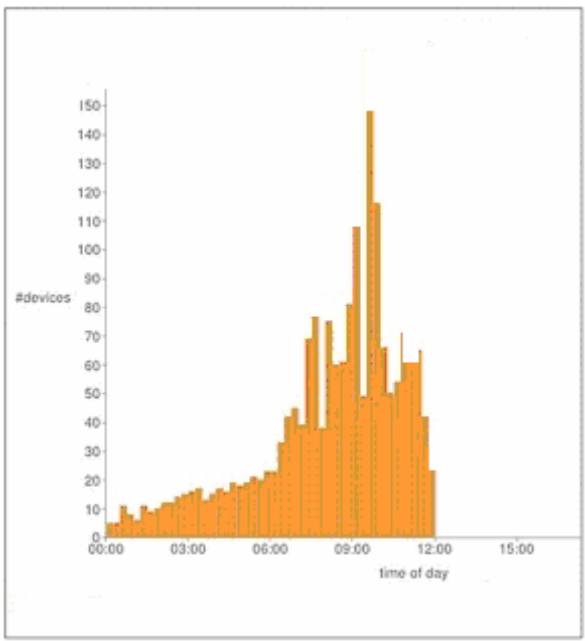

Fig.4. Location analytics (by Cisco MSE)

Also, we can try to detect the time visitors spend at the given location. But it depends on the phone's ability to send several Probe Requests within the investigated time period. In other words, it is not 100% reliable measurement.

More interesting data could be obtained with several registering devices within the area. MAC-address plays the role of unique cookie in web analytics. As soon as we get two sequential records for the same MAC-address, we can make the assumption about the direction (route). Also, we can estimate the speed and moving patterns (it is more interesting for 3 and more registering devices).

Figure 5 illustrates the standard location analytics report by Navizon. In general, it could be described as a real analytics for the real places. It is what makes Google Analytics for web sites, but applied for the real places and real visitors. Let us see, for example, the standard set of reports from Google Analytics: Overview, Demographics, Behavior, Technology, Social, Mobile, and Visitors Flow. Of course, we cannot map them one by one to the new analytics, but the basic elements could be reproduced. For example, the Overview report could be presented directly, Demographic could be predicted (like search engines do demographic predictions by visits and clicks), Behavior could be obtained from several passive Wi-Fi monitoring centers, passive monitoring can get vendors info for mobile phones (Technology), etc. Note also, that Google Analytics can provide real time data and what is especially interested for our development – API level.

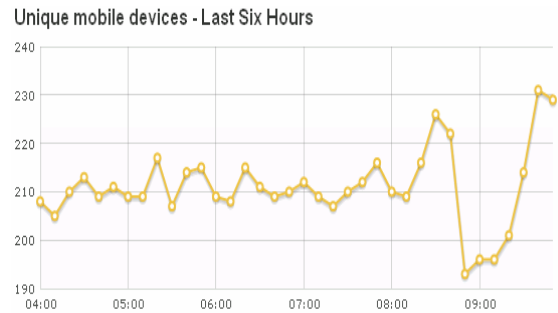

Fig.5. Unique visitors (by Navizon I.T.S.)

In this paper we propose a new model (use case) for passive monitoring. It is messaging for the real places and real visitors.

II. CLOUD MESSAGING

Google Cloud Messaging for Android (GCM) is a service that allows you to send data from your server to your users' Android-powered device. This could be a lightweight message telling your app there is new data to be fetched from the server (for instance, a movie uploaded by a friend), or it could be a message containing up to 4kb of payload data (so apps like instant messaging can consume the message directly).

The GCM service handles all aspects of queuing of messages and delivery to the target Android application running on the target device. GCM is completely free no matter how big your messaging needs are, and there are no quotas [13].

There are conceptually similar services from other wendors (e.g., Apple, Microsoft, Nokia). Architectures of these push notification services have common features. At the first hand, application servers send a notification message with an intended receiver (or the target mobile device) to one of the cloud-based messaging servers. Messaging servers pushes the message to the target mobile device. The push notification service eliminates the needs of application servers to keep track of the state of a mobile device (i.e., online or offline). Furthermore, mobile devices do not need to periodically probe (poll) the application servers for messages. It reduces the workloads of the application servers and seriously simplifies the mobile application development.

We describe below Google Cloud Messaging Service as a main system used in our development. In the same time principles are the same for all the above-mentioned services.

Here are the primary characteristics of GCM as per Google's manual. GCM allows 3rd-party application servers to send messages to their Android applications.

An Android application on an Android device doesn't need to be running to receive messages. The system will wake up the Android application via Intent broadcast when the message arrives, as long as the application is set up with the proper broadcast receiver and permissions.

The first time the Android application needs to use the messaging service, it fires off a registration Intent to a GCM server. This registration Intent includes the sender ID, and the Android application ID.

If the registration is successful, the GCM server broadcasts an intent which gives the Android application a registration ID.

The Android application should store this ID for later use. To complete the registration, the Android application sends the registration ID to the application server. The application server typically stores the registration ID in a database.

The registration ID lasts until the Android application explicitly un-registers itself, or until Google refreshes the registration ID for your Android application. Figure 6 illustrates the whole process.

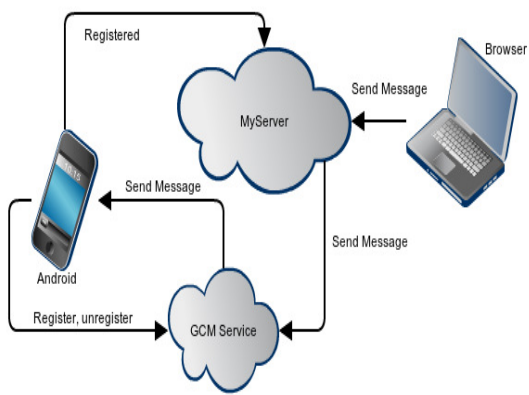

Fig.6. GCM workflow

For an application server to send a message to an Android application, the following things must be in place:

- The Android application has a registration ID that allows it to receive messages for a particular device.

- The 3rd-party application server has stored the registration ID.

- An API key. This is something that the developer must have already set up on the application server for the Android application. Now it will get used to send messages to the device.

Here is the sequence of events that occurs when the application server sends a message:

- The application server sends a message to GCM servers.

- Google en-queues and stores the message in case the device is offline.

- When the device is online, Google sends the message to the device.

On the device, the system broadcasts the message to the specified Android application via Intent broadcast with proper permissions, so that only the targeted Android application gets the message. This wakes the Android application up. The Android application does not need to be running beforehand to receive the message.

Any Android application cans un-register GCM, if it no longer wants to receive messages [14].

### III. LOCAL MESSAGING

Based on the above-mentioned description, we can note that receiving the messages requires the registration phase. Android application (read – mobile phone with installed application) should inform GCM about the possibility to obtain messages. Usually, this contract is presented in the form of some ID (registration ID). IDs are stored in database. So, our service can select all the stored IDs and distribute some custom message (messages) to applications.

What if we include into process of registration MAC-address too? This decision lets us simply compare subscription info with the locally detected (presented) mobile subscribers. Of course, the mobile application for the subscription can read MAC-address automatically. The whole schema for the service becomes very transparent.

1) We can divide our data by topics.

2) Each topic corresponds to some location with passive Wi-Fi monitoring equipment. For example, we can create a topic, which corresponds to some café, or a topic, which corresponds to some building in the University campus, etc.

3) There is an administrative software that lets authorized people (topic's admin) add (edit, delete) some messages to topics. For example, it is a message, which describes new discount in café, or it is a message, which informs about new seminar in our university, etc.

4) Mobile user informs our system about his intention to receive messages from some topic (theme). This registration procedure will include GCM registration too. So, for every new subscriber our messaging system will save his topic ID (IDs), GCM ID and MAC-address in our own database.

5) Wi-Fi monitoring detects the presence for mobile phones. Actually, as any mobile phone has been described by its MAC-address, we will detect the presence of our subscribers.

6) Our daemon scans the detection log, extracts MAC-addresses and compares them with subscription database for this location.

7) As soon as we discover that our subscriber is detected (he is somewhere nearby), we can use CGM for delivering the above-mentioned custom messages.

Note, that MAC-address in this schema is used for the re-identification only. So, for keeping the privacy, we can replace it with some hash-code (for both processes: monitoring and subscription).

Let us describe the possible schemas for the delivery of our messages.

In the simplest case, we can deliver each message to the every locally detected subscriber. More sophisticated conditional delivery terms are presented below.

- Deliver some messages several times with the predefined intervals between attempts.

- Delivery messages at the pre-definite time only. E.g., we can define, that the particular message should be delivered from 2 p.m. till 3 p.m. only.

- Lets mobile users temporarily switch off/on the subscription

- Add customized rules for sending

Rule-based delivery is the most interesting part. Practically, we implemented here the same approach as in the above-mentioned SpotEx system [5]. Our messaging server presents rule-based expert system. As usually, each rule (production) has got condition (left part) and conclusion (right part). The conclusions are standard. Each our conclusion just states, that an appropriate message should be delivered (pushed) to the mobile terminal. Each condition, in the same time, could be presented as a logical expression with build-in functions. At this moment we can list the following standard functions:

COUNTER( )

FIRST ( )

IN_PLACE ( )

IN_GROUP_OF ( )

SUBSCRIBED_TO ( )

Function COUNTER (n) returns a number of visits. Its argument describes a time interval. Possible values for time interval are:

0 - all time

1 – a day

2 – a week

3 – a month

E.g., CONTER(2) call returns the total number of visits for the current week.

Function FIRST(n) returns a boolean value *true* if the visit is first one for the given time interval.

Function IN_PLACE(t) returns a boolean value true if visitor stays in place at least t minutes. Note, that due to the passive monitoring principles it could not be detected for all users (it depends on probe requests from the mobile phone).

Function IN_GROUP_OF(n) returns a boolean value true if server detects at least *n* visitors at the moment of calculation.

Function SUBSCRIBED_TO(id) returns a boolean value true if user (mobile phone) is subscribed to the given topic (*id* describes a topic). For example:

IF COUNTER(3)>2 AND FIRST(2) THEN

{ *deliver coupon info message* }

Coupon info should be delivered if user (mobile phone) has 2 or more visits per month and current visit is the first visit for this (current) week.

It other words, we can describe this implementation as a message-oriented version for SpotEx. Similar to the SpotEx model, our rules present the standard production rule based system, and we can use Rete algorithm for the processing. A Rete-based expert system creates a network of nodes. Each node (except the root) corresponds to a pattern presented in the left-hand-side (in the condition) of a rule. The path from the root node to a leaf node defines a complete rule's condition. Each node has a memory of facts, which satisfy that pattern. This structure presents essentially a generalized tree. As new facts are asserted or modified, they propagate along the network of nodes. It causes nodes to be annotated when new fact matches existing pattern. When a fact or combination of facts causes all of the patterns for a given rule to be satisfied, a leaf node is reached. Leaf node triggers the rule [1]

The typical use cases are proximity marketing and news delivery in Smart City projects for example.

The push notification services on other platforms are similar to GCM in the architectural design. When an application launches in a mobile device, it needs to register to the push service to get a unique ID. This ID may have different names in different platforms, e.g., device token in iOS and push URI in Windows. After that, device sends ID (token, URI) to the application server. When the application server wants to send a push notification to an application, it sends the ID together with the payload to a push server. Push server forwards the payload to the application [15].

What are the advantages for this approach? At the first hand, it is so-called passive monitoring. There are no special applications for mobile subscribers (except subscription/unsubscription service). The messaging will target only subscribers physically presented in the covered area. The process for subscription and un-subscription is very straightforward. The "check-in" process (passive discovering) is secure. It does not keep records in social networks like ordinary check-ins in Foursquare, Facebook, etc. It does not require user's identification too. It works as an extension for customized check-ins [16].

What are the disadvantages? At the first hand, the passive monitoring (as we wrote above) is not 100% reliable. Push messaging delivery requires internet connectivity. But in the same time, installing active Wi-Fi access point on-site, mobile users can connect to, will improve the discovery process. Also, we can use SMS as a backup channel (if phone number is provided during the subscription, of course).

IV. CONCLUSION

This article presents a new mashup based on passive Wi-Fi monitoring for mobile devices and cloud based notifications. Passive monitoring uses probe requests from Wi-Fi specifications for detecting nearby clients. Notification module uses cloud messaging (push notifications) from mobile operational systems. This approach does not require special mobile applications for mobile users. This application does not publish location info in the social network. Practical use cases for this application are proximity marketing and Smart City projects. The proposed approach automatically guaranties that custom messages will target online subscribers in the nearby area only.

ACKNOWLEDGMENT

The paper is financed from ERDF's project SATTEH (No. 2010/0189/2DP/2.1.1.2.0/10/APIA/VIAA/019) being implemented in Engineering Research Institute "Ventspils International Radio Astronomy Centre" of Ventspils University College (VIRAC).